\documentclass[conference]{IEEEtran}
\pdfoutput=1

\usepackage{pdfpages}
\usepackage{hyperref}
\usepackage[absolute]{textpos}
\IEEEoverridecommandlockouts
\usepackage{makecell}
\usepackage{xcolor}

\usepackage{wrapfig}
\usepackage{graphicx}
\usepackage{cite,url}
\usepackage{array,multirow}
\usepackage{tabularx,boldline}
\usepackage{cellspace}
\usepackage{balance}

\usepackage{expdlist}

\usepackage{amsmath,amssymb,amsmath,amsfonts}

\def\BibTeX{{\rm B\kern-.05em{\sc i\kern-.025em b}\kern-.08em
 T\kern-.1667em\lower.7ex\hbox{E}\kern-.125emX}}

\begin{document}

\begin{textblock}{30}(1,0.2)
\noindent\tiny  This paper is a preprint; it has been accepted for publication in 2020 IEEE World Congress on Services (SERVICES), 18-23 Oct.  2020, Beijing, China.\\
\textbf{IEEE copyright notice} \textcopyright 2020 IEEE. Personal use of this material is permitted. Permission from IEEE must be obtained for all other uses, in any current or future media, including reprinting/republishing this material for advertising or promotional purposes,\\ creating new collective works, for resale or redistribution to servers or lists, or reuse of any copyrighted component of this work in other works.
\end{textblock}
\bstctlcite{IEEEexample:BSTcontrol}

\bstctlcite{IEEEexample:BSTcontrol}
\balance
\title{On the Security and Privacy of Hyperledger Fabric: Challenges and Open Issues %
\thanks{%
\protect\begin{wrapfigure}[3]{l}{.9cm}%
\protect\raisebox{-12.5pt}[0pt][0pt]{\protect\includegraphics[height=.8cm]{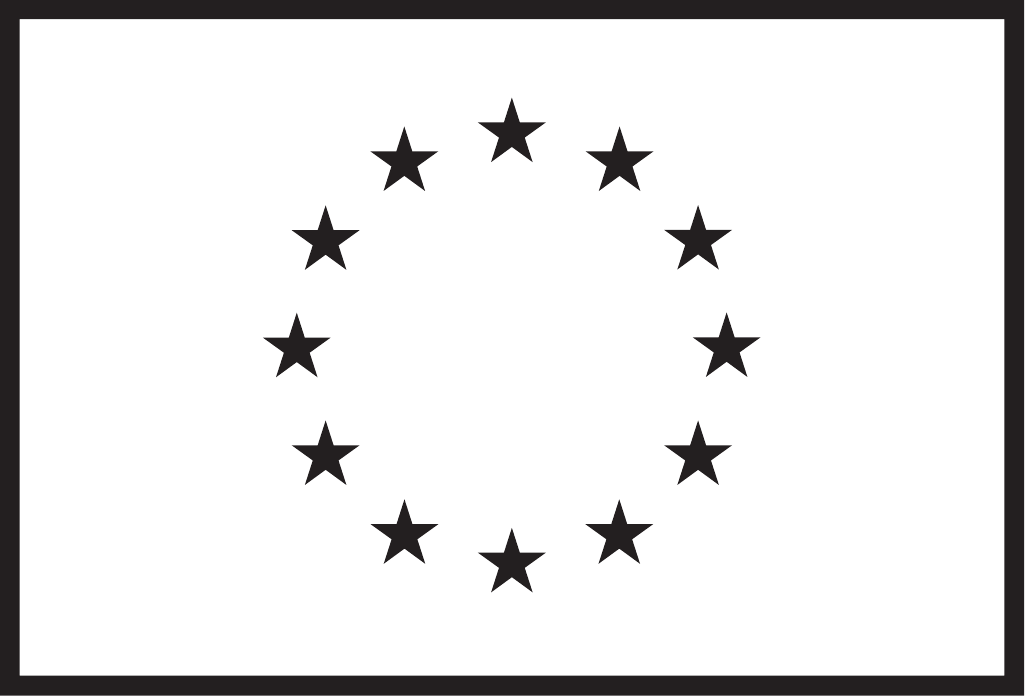}}%
\protect\end{wrapfigure}%
This project has received funding from the European Union's Horizon 2020 research and innovation programme under grant agreement no. 786698. The work reflects only the authors' view and the Agency is not responsible for any use that may be made of the information it contains.}}

\author{\IEEEauthorblockN{%
Sotirios Brotsis\IEEEauthorrefmark{1},
Nicholas Kolokotronis\IEEEauthorrefmark{1},
Konstantinos Limniotis\IEEEauthorrefmark{1},
Gueltoum Bendiab\IEEEauthorrefmark{2}, and
Stavros Shiaeles\IEEEauthorrefmark{2}
\vspace*{4pt}}
\IEEEauthorblockA{\IEEEauthorrefmark{1}University of Peloponnese, 22131 Tripolis, Greece\\
Email: \{brotsis,\,nkolok,\,klimn\}@uop.gr}
\IEEEauthorblockA{\IEEEauthorrefmark{2}University of Portsmouth, PO1 2UP, Portsmouth, UK\\
Email: \{gueltoum.bendiab,\,stavros.shiaeles\}@port.ac.uk}
}

\maketitle

\begin{abstract}
In the last few years, a countless number of permissioned blockchain solutions have been proposed, with each one to claim that it revolutionizes the way of the transaction processing along with the security and privacy preserving mechanisms that it provides. Hyperledger Fabric is one of the most popular permissioned blockchain architectures that has made a significant impact on the market. However, there are only few papers of finding architectural risks regarding the security and the privacy preserving mechanisms of Hyperledger Fabric. This paper separates the attack surface of the blockchain platform into four components, namely, consensus, chaincode, network and privacy preserving mechanisms, in all of which an attacker (from inside or outside the network) can exploit the platform's design and gain access to or misuse the network. In addition, we  highlight the appropriate counter-measures  that can be taken in each component to address the corresponding risks and provide a significantly secure and enhanced privacy preserving Fabric network. We hope that by bringing this paper into light, we can aid developers to avoid security flaws and implementations that can be exploited by attackers but also to motivate further research to harden the platform's security and the client's privacy.
\end{abstract}



\section{Introduction}
\label{sec.intro}
Hyperledger Fabric \cite{Androulaki} (for simplicity Fabric), has recently obtained massive popularity with hundreds of implementations  around the world, since it is quite scalable, lenient against faults, and robust. For these reasons, among others,  it can satisfy more than enough  and sufficiently better than any other permissioned blockchain solution  the purposes of  an enterprise-based environment. Only in the past year, it is considered as the most deployable distributed ledger, in various areas of interest, such as the IoT ecosystem \cite{Brotsis}, the supply chain finance \cite{Ma}, the medical data management \cite{Medical} and more. 

By its permissioned nature, Fabric  is a closed system, in which only the participants that have obtained the necessary credentials are able to read or write to the ledger. These participants are called peers and only a subset of them can approve transactions and in order to do so, they have to mention  their  identity  along  with  their  signatures \cite{Kolokotronis}. This setting makes it easier for the peers to manage the transactions on the ledger and it is typically the reason why Fabric is much faster than the ongoing permissioned blockchains.  

For the maintenance of the identities of all the  participating nodes (clients,  \textit{ordering service nodes} (OSNs) and  peers) responsible is the membership service provider, which is one the most critical components of the platform,  since it manages any type of access by issuing  credentials in a form of cryptographic certificates that are used for authentication and authorization. 

The features of Fabric are not limited only to its design, the support of pluggable consensus, which is another critical component; provides an  unprecedented level of extensibility and specific the support of multiple ordering nodes that  establish consensus regarding the transactions' total order. Moreover, since version $1.0$, Fabric's ordering service comes without any \textit{Byzantine Fault Tolerant} (BFT) consensus protocol that can address possible malicious ordering nodes, implementing only \textit{Crash-Fault-Tolerant} (CFT) protocols based on Kafka and on Raft. 

Nearly almost all the permissioned blockchain solutions can implement smart contracts, which are based on a  programmable application logic that is being called each time a transaction is being proposed. In Fabric's case the smart contracts are realized by means of  an arbitrary program that is authored in Go; the chaincode. The chaincode  is executed by a set of peers locally and before each transaction is appended into the ledger, an output of the chaincode's execution is taken into account in order to decide whether a transactions is valid or not and which data will be included to the ledger. 

\begin{figure*}[ht]
\centering
\includegraphics[width=\textwidth]{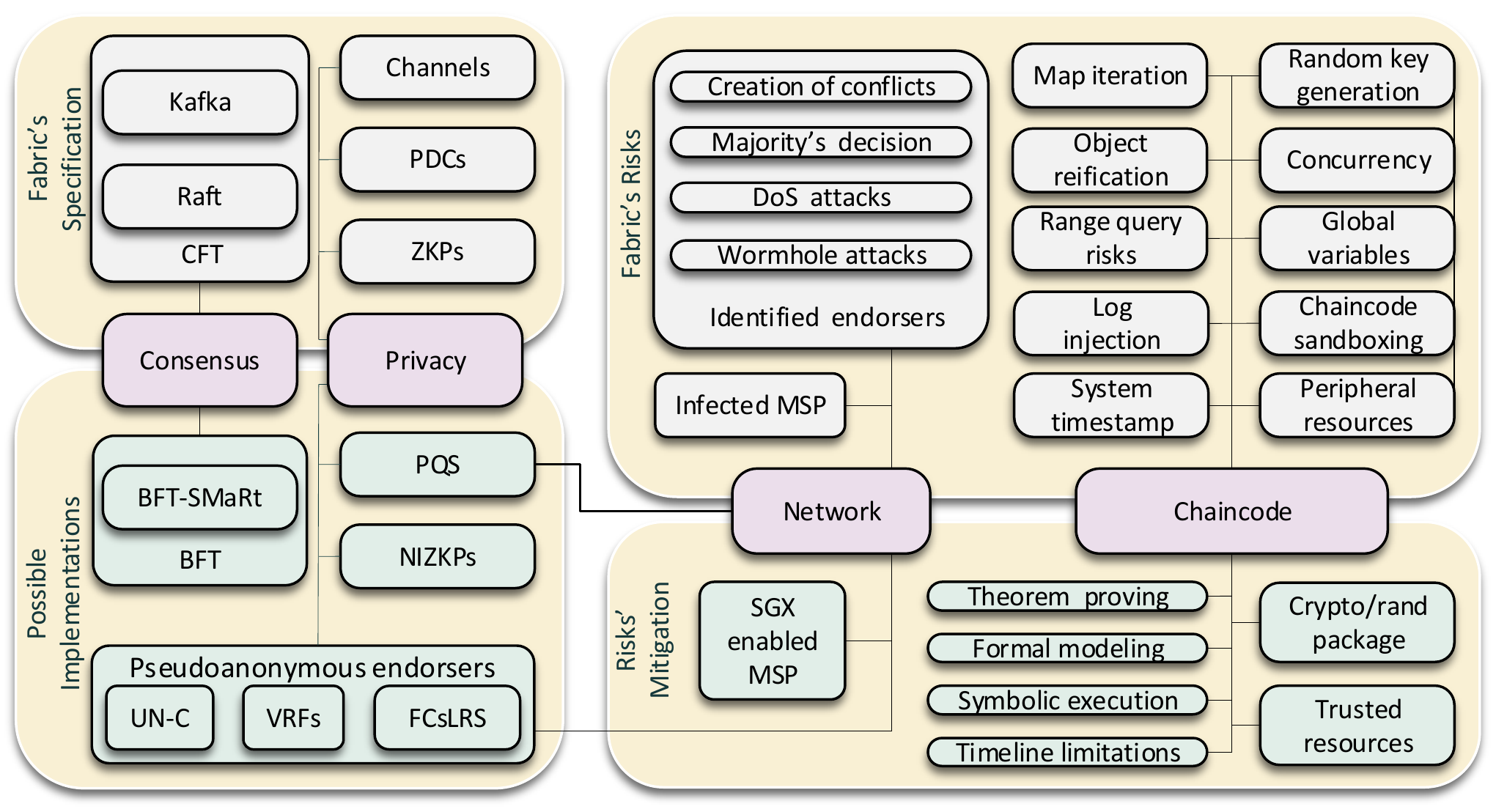}
\caption{Hyperledger Fabric's security architecture}
\label{fig.sec}
\end{figure*}

From the security perspective,  we analyze  Fabric into four interconnected components, in which possible attacks and leakage of private information can occur;  namely: the consensus, the chaincode, the network  and its privacy preservation mechanisms. As a result of our research, in each of these components, we define the  risks and we provide the corresponding counter measures needed to be addressed and enhance its security: $a)$ The implemented consensus protocols in Fabric can withstand only some of the ordering nodes to crash but not to behave maliciously. $b)$ The privacy protection features of  Fabric include the following four aspects, namely the channels, the private data collections, the zero-knowledge proofs and the membership service provider, but the platform does not support any non-interactive zero-knowledge proofs and post quantum signatures. $c)$ and $d)$ Regarding the security of the smart contracts (and the network's) we have  surveyed and outlined the potential vulnerabilities associated with Fabric's chaincode (and the network's layer) and provided the corresponding countermeasures and verification methods for these vulnerabilities to be averted, as shown in Figure \ref{fig.sec}. To the best of our knowledge there is no other published research that has assembled from all the four components,  the possible threats  that can occur in Fabric and has discussed possible counter-measures to mitigate these threats.

The  remainder  of  this  paper  is  organized  as  follows. Section \ref{sec.Security from Consensus} presents the security that Fabric provides from the consensus process and a BFT consensus protocol that can be implemented in the presence of malicious OSNs along with a comparison of these protocols.  Section \ref{Security from the smart contracts} surveys a number of smart contracts' vulnerabilities and the possible counter measures and methods that can be taken to enhance and verify the chaincode's security. In Section \ref{Network's security} we analyze the impact to the network; if the membership service provider is compromised, the possible attacks that can occur from having the endorsing peers been identified and we propose possible countermeasures. In Section \ref{Privacy Preservation}, we analyze how Fabric satisfies many privacy flavors that and we discuss the open challenges to these issues. 

\section{Consensus security}
\label{sec.Security from Consensus}

Consensus in blockchain technology is a state, in which all the participants agree upon an identical sequence of the messages in a specific order \cite{Androulaki}, i.e. the atomic broadcast\cite{Cristian}. In Fabric the OSNs atomically broadcast endorsements to establish agreement, by means of consensus, regarding the transactions' total order. 

\subsection{Consensus methods and Fabric's specification}
The support of pluggable consensus  protocols is a valuable discrimination in each blockchain protocol to achieve a trust model or other use cases \cite{Androulaki}. Assuming that a blockchain protocol is deployed in a small environment, maintained by a small enterprise or a single trusted-third party; the implementation of complicated  protocols  can be seen as an exaggerated drag on the energy's consumption. For this reason, the adoption of a single OSN to achieve consensus and order the clients' proposals in a small and highly trusted environment can be proved to be a  quite as good choice as the adoption of other perplexing consensus protocols.  

\subsubsection{Solo}

As a centralized consensus protocol, Solo is being implemented for testing purposes, since it necessitates only a single node to receive and order the incoming transactions in order to provide an inferior consensus method. By adopting this  approach, developers can focus on other concerns such as the development and improvement of the chaincode. Nevertheless, Solo's outcome is the creation of a \textit{Single-Point-of-Failure} (SPoF) and if this single ordering node crashes the entire system will collapse. 
 
\subsubsection{Apache Kafka}

Kafka \cite{kafka} is a  distributed publish/subscribe messaging pattern that is being used to transfer large amount of log data with significantly low latency. Kafka's key components are the producers, the topics,  the consumers and the brokers. The recorded information is published from the producers to a stream of messages called the ``topic", which is a partition of segments of files. The messages are stored from the brokers as the latest segment file and when the producers publish messages to the  partitioned logs, only the subscribed consumers can  consume these messages - sequentially - by making requests to the brokers. Kafka's fault tolerance properties  derive from the ZooKeeper \cite{ZooKeeper} and they can be achieved with the replication of the partitions between the brokers. Since, Kafka follows the leader-follower design, each partition has its own leader, whose actions are passively replicated by its followers. Therefore, if the ongoing leader  crashes, a new election process begins with one of its followers to take its place.  Regarding its performance,  Kafka showed momentous results in \cite{Androulaki} and brought significant enhancements in the area of the business oriented and permissioned  blockchains. Thus, Kafka, surpassing Solo, is the suggested protocol in  version $1.0$.  

\subsubsection{Raft}
\label{Raft}

The Raft consensus protocol \cite{Raft}  is also based on the “leader-election” model to  establish consensus by electing a leading node to acquire the incoming entries from  the clients and replicate them. To provide strong leader and coherency, the protocol is separated into three phases, i.e. the leader election, the log replication and the safety.  The time in Raft proceeds in  arbitrary time periods, called ``terms", with each term to be defined by an increasing number. The nodes in Raft are hierarchically ranked in different states, with each node to be either a leader, a follower or a candidate. The  leader is the principal entity of the protocol and it  is elected per channel, with the task to interact  with the clients and then replicate its entries to its synchronized followers. Therefore, in order to achieve the best possible synchronization, it sends systematic heartbeats to its followers and even if the network suspects that the leader has crashed, at least one of its followers will detect this  divergence, cast a vote to the network and attempt to take its place \cite{Bellini2}. Some nodes might compete to win the election by seeking votes from other nodes. Therefore, these nodes are considered as ``candidates''.  

Raft ensures that exclusively a single node can become a leader, even if some nodes miss a term or split ups occur. In the first case the outdated node will revise the term's number and fall into the follower's state; and in the latter, the current term will be ended without any outcome of the election process. Raft's performance is not fully tested, but its adoption  as the recommended consensus protocol since the version $1.4.1$, showed that  Raft can provide thousands of transactions in real world scenarios \cite{Fabricdocs} and twice the transaction's throughput of Kafka with even less latency \cite{raftperformance}. 

\subsubsection{BFT-SMaRt}
\label{BFT-SMaRt}

BFT-SMaRt \cite{BFT-SMART} is java-based consensus protocol that can provide a secure ordering service for Fabric \cite{Sousa}. In the ideal case where no adversarial \textit{validation replicas} (VRs) exist in the network, the BFT-SMaRt’s message processing is identical to  PBFT’s \cite{Pbft}.  
 
 WHEAT is the component that BFT-SMaRt's  ordering service \cite{Sousa}  relies upon \cite{WHEAT} to provide  a powerful vote assignment scheme, low latency and fast replication among the   VRs, without imperiling the network's security. The ordering service consists of the cluster nodes and the frontends. By adopting BFT-SMaRt as the consensus protocol, the transaction flow is almost the same to \cite{Androulaki}.  Upon the collections of  the endorsements from the peers, a client generates a signed envelope that contains  the channel's identity  and the peer's endorsements accompanied by their signatures \cite{Androulaki}. This envelope is  disseminated to the frontends and finally sent to the OSNs for ordering. When the OSNs collect a predetermined number of envelops from some trusted frontends or when a predetermined time has elapsed,  a new block is created that contains only the valid transactions. Therefore, the transactions' validation differs from \cite{Androulaki} and it occurs prior to the creation of the block's signature.  Finally, all the same, the block is transferred to the frontends and then to the peers that manage the ledger. BFT-SMaRt's performance is measured in \cite{Sousa} and showed that it can  achieved  throughput more than $10$ thousand TPS and the block confirmation time to be less than $1$sec. 

\subsection{Comparison}

The adoption of Solo creates a SPoF, since it is not lenient against malicious failures and crashes and for this reason, it should not be deployed in real world environments. Both Raft and Kafka benefit from the ``leader-follower'' model to address crashes, but  despite Kafka's popularity, various intricate components have to be handled for its implementation. In Raft's case, these components are enclosed to the ordering service \cite{Fabricdocs}, meaning that there are less components that might crash. Kafka is designed to be deployed in an environment with a small number of OSNs and the  cluster to be managed by a single entity. This concept does not contribute much to escalate the system's decentralization,  forcing all the OSNs to be ulcerated by a single entity.  In contrast to Kafka, a Raft-based ordering service is more decentralized, more scalable and it can achieve a greater throughput \cite{raftperformance}. For reasons as such, the Kafka and the Solo consensus protocols are deprecated in Fabric's  version $2.0$.  Despite Kafka's deprecation though, a  reconstruction of the Fabric's transaction flow managed to increase Kafka's throughput from $3.5$ to $20$ thousand TPS \cite{FastFabric}. 

On the other hand, BFT-SMaRt can not only achieve a remarkable transaction performance of $10$ thousand TPS but it also can withstand tolerance against possible malicious behavior by the OSNs. In BFT-SMaRt, the invalid transactions are not included in the blocks, since the transactions' validation occurs prior to the block's creation and dissemination to the peers.  Although  BFT-SMaRt seems to outperform Raft, both in terms of security and  performance; it is a Java-based library that it does not provide a  very stable ordering service  and therefore it is not being currently adopted by the Hyperledgr Project. BFT-SMaRt's drawback though is that  it necessitates  two processes to operate; the first built in Java and the second from GO, using a network socket bound among them, which can act as bottleneck in the system's operation. 

\subsection{Consensus' challenges and open issues}

The consensus protocol is the most critical component of a distributed ledger. CFT protocols are consider  to be contradicting to the platform's security, since any malicious action can affect the network's security. The BFT protocols are  adopted to permissioned blockchain solutions for targeted use cases, in which the requirements to provide a secure implementation is more evident.  The fact that each peer in a permissioned network is accountable for its behavior, provides an incentive to the nodes to adhere to the protocol. In spite of the tolerance that the consensus protocol provides (i.e. CFT or BFT), Fabric can mitigate the most familiar and sophisticated consensus-oriented attacks, such as the double spending, by its design.  Therefore, the consolidation of Fabric and a CFT consensus protocol can be considered as an ideal implementation in a  confidential network, such as an  enterprise environment.  

\section{Smart contracts' security} 
\label{Security from the smart contracts} 
The execution of the smart contracts defines operations that have been acknowledged by all the  participating entities; with each one to execute the contract locally,  propose the result to the network and then  collaborate with it to select which result is going to be inserted into the ledger. The development of smart contracts can incorporate several programming errors  that can ultimately lead to exploitable bugs and malicious behavior. Consequently, the  smart contracts are prone to code errors and inconspicuous vulnerabilities, while their accuracy and security can be violated by malicious programmers by means of  exploits. 

\subsection{Chaincode' vulnerabilities and Fabric's specification}
\label{Potential vulnerabilities}
The programs’ defects, such as coding flaws and designing errors, are the main reasons that cause the smart contracts’ vulnerabilities. In \cite{Atzei}, a large variety of Ethereum smart contracts vulnerabilities have been identified and showed that programming flaws lead to  faulty  behavior, a series of attacks and possible exploits. As we are focused on Hyperledger Fabric and particularly to the Fabric's chaincode,  our following discussion is focused on risks that might ascend from the programming languages, the Fabric's features and the  mis-understanding of the common practices. 

Fabric's smart contracts were not built with the vision of being strained to a \textit{domain-specific programming language} (DSL), such as the Ethereum's Solidity, but rather to be authored in high-level languages (such as Go, Java and Node.js)   to reduce the developers' learning cost.  Formally, the smart contracts that have been authored in DSLs are ruled by particular features and  restrains for blockchains. Since this is not the case of Fabric,  the known risks and vulnerabilities might  differ with the risks associated with general-purpose programming languages.  Therefore, based on \cite{Yamashita},  \cite{Hasanova}, \cite{Tobias} and \cite{Huang}, we outline in the following table the most prominent security vulnerabilities in Fabric's chaincode. The programming language Go is the most widely used  during the chaincode's development and thus, most of these security vulnerabilities derive from its non-deterministic behavior. Nevertheless, Fabric has no native cryptocurrency built in and thus, it is not easy to define how severe each vulnerability is and how easy or difficult it is for an attacker to exploit possible bugs and execute double spending attacks.

In the execution phase, the peers do not execute the chaincode at  the same time andin the same environment. If a transaction is valid and the results that  derive from the chaincode's   execution   are  not deterministic,  in the  validation phase these results might be rejected or allow double spent transactions to be included to the ledger. The risks that derive from the Go language, from the platform's features or from mis-understanding of the common practices can lead to inconsistencies to the peers' ledger: 

\subsubsection{Random key generation} Since, the simulation of the chaincode occurs in each endorsing peer, the random seed for the keys is different. In  Go, the random seed is set as $1$  and therefore can be easily predicted by a client.

\subsubsection{Object reification} The value of the variables are handled through a pointer, which is an address of memory. Therefore, using reified object addresses might cause non-determinism.

\subsubsection{System Timestamp} It is difficult to ensure that the timestamps are run concurrently in each peer.

\subsubsection{The global state  variables}   Global state variables that are not stored to the ledger, might change  innately  and  cause inconsistencies to the endorsing results.

\subsubsection{Concurrency} If multiple transactions are executed concurrently  and under high load, a possible change at the keys' versions might lead to key collisions and double spending. For example, if a transaction has passed from the endorsing phase and its version key has changed before it reaches the validation phase not only a program error will occur but this action might also allow another (possible double spent) transaction to be included to the ledger.

\subsubsection{The non-determinism that ascends  from peripheral resources} Some accessing resources, such as web services, external libraries, external  files and system command executions;  can corrupt the code and return different endorsing results among the endorsing peers.

\subsubsection{Range query risks} Queries methods to access the Fabric's state databases and obtain private data, (e.g. the history or the state of a key); are not executed again in the validation phase and can lead to phantom reads, in which the dirty data cannot be detected.

\subsubsection{Chaincode sandboxing} Although, Fabric's chaincode is executed in an isolated docker  container and provides  sufficient privileges, it  can be  exploited in a malicious way, such as to execute port scanning, identify and exploit vulnerable peers, install malicious software and execute attacks. 
\subsubsection{Log injection} Any corruption of the log messages  can possible avert them from being executed automatically and allow the attacker to view the processed logs.  
\subsubsection{Map structure iteration} Due to the hidden implementation details of the Go programming language, when an iteration with map structure is used, non-determinism may arise and the order of key values might be different.

\subsection{Chaincode's challenges and open issues} 
A docker container, by its design, is significantly secure, particularly when the processes are executed with no privileged users in the container. Albeit, developers should be familiar with the dockers' security issues to fully take advantage of the security and the efficiency that the dockers  provide. Thus, in this section we discuss the proactive measures that can be reserved to increase the smart contracts' security:

\begin{itemize}
\item Timeline limitations of the chaincode's execution can be employed;   
\item The chaincode's execution can be performed with privileges other than root;  
\item  GO's crypto/rand package can be used to produce  a crytographically  secure random seed for the keys and minimize the risk of double spending; 
\item  Any changes at the keys' versions can increase the possibility of  double spending and thus,  should be avoided;
\item  Only trusted services peripheral of the platform should be accessed; 
\item Chaincode’s input arguments should be checked for escape characters; 
\end{itemize}

The aforementioned counter-measures can be easily taken into account during the chaincode's implementation. Nonetheless, a number of other techniques \cite{Hasanova} are relied upon  formal verification approaches  to mitigate risks in smart contracts:

\subsubsection{Theorem proving} One of the most ordinary methods to formalize smart contracts. Symbolic logic (with axioms or premises) is used in theorem proving  to prove the necessary security properties that enhance the contracts'   accuracy.  

\subsubsection{Symbolic execution tools} This  technique uses testing tools, such as the Chaincode Scanner,  to perform static analysis and find bugs, vulnerabilities and bad practices in the chaincode, as well as, to provide a detailed description of a possible  problem. 

\subsubsection{Formal modeling} This method   formalizes the smart contracts' risks, by using precise statements that define the relationship between the smart contracts'  components resulting in $a)$  unambiguous communication, and $b)$ replicable/reproducible results.  These risks are classified  into seven categories \cite{Singh}, that analyze the platform's properties; specifically:  ``privacy,  security, bug bounty, trustworthiness of data, scalability and  correctness of the consensus protocol. 

Although, these techniques are not the most common for verification of the smart contracts (according to \cite{Singh}), they seems to be a quite promising field of research, especially in  the Fabric's case \cite{Ardagna}.   

\section{Network's security}
\label{Network's security}
With the increase of interest in permissioned blockchains; networks that possess the qualifications to handle administrative permissions, the possibility of an anonymous attacker  feigning to behave correctly is almost minimized. Attacks and strategies such as the $51\%$, the Sybil attack, the block withholding and the selfish strategy are lesser risks to the network's security \cite{Kolokotronis2}, \cite{Shala}, due to the trust and the restricted permissions to those given access to the network.  

\subsection{Network's threats and Fabric's specification}
\label{Potential risks}
The rise of Fabric, on the other hand, comes with new security risks and concerns that can harm the network's operation and performance \cite{Davenport}.  Therefore, in this section, we present the most prominent attacks that derive from the network's level, if some participating entities are compromised; along with feasible countermeasures to enhance its security. 

\subsubsection{Compromised membership service provider (MSP)}
The most revolutionary aspect of blockchains lies in the decentralized nature of a trustless network and Fabric seems to  violate just that. The centralized aspect of Fabric lies on the MSP and the following \textit{Certificate Authority} (CA). The  MSP is a critical component of the platform, since it manages the registration, the identities and the type of access of all the nodes in the network; comprising the clients, the peers and the OSNs.  Therefore, if the MSP is compromised, administrative controls such as adding and removing identities to and from the network, as well ass, the type and the amount of the given  access to the existing nodes are all managed entirely by the attacker. With a malicious MSP, the unauthorized access that is given  to the attacker can cause serious damage and possible lead to further attacks such as: invalid identification attack,  double-spending,   attacks on the CA, etc. \cite{Davenport}.  

\subsubsection{Identified endorsers}
\label{Identified Endorsers}
In Fabric's execution phase, a transaction needs to be approved by a set of endorsers. In return, the endorsers,  approve the transaction mentioning their identity along with their signatures so that they can be verified in the forthcoming phases. However, Fabric has some drawbacks when the peers are identified \cite{Andola}, \cite{Mazumdar}, namely:
\begin{itemize}
\item Creation of conflicts: For some transaction, the endorsing peers may have different opinion about their validity. By revealing their identities, possible  conflicts can be created within the consortium \cite{Mazumdar}.

\item  Majority's decision: The endorsement policy not only  does not allow the endorsers to approve the transaction in secret, but also takes into consideration the majority's decision. 

\item  DoS attacks: An opening for DoS attacks on selective endorsers might be created, either to halt specific transactions from being included to the ledger or to degrade the network's performance \cite{Andola}.  

\item  Wormhole attacks: An opening for  wormhole attacks might be created, if a peer  behaves maliciously and colludes with an adversary exterior to the channel. This attack can lead to leakage of private information of all the peers of a specific channel. This problem is concluded by defining the fact that the Fabric's access control mechanism  is depended on trusting each peer inside a channel \cite{Bellini:}, \cite{Bellini3}. 
\end{itemize}

\subsection{Network's challenges and open issues}
\label{Network's Challenges  -  Open  Issue}
The aforementioned analysis for the MSP and the endorsers indicates that the appropriate safeguards are not in place to mitigate network's risks in the context of Fabric. Although the platform offers strong accountability in the network to meet  the necessary flavors of security, the research in this field is still ongoing. Thus, some other possible approaches that could be appropriately adopted in the (near) future are:

\subsubsection{Securing the MSP} 
The possible threats that can derive from a compromised MSP are tackled in a recent study \cite{Liang} with the adoption of \textit{Intel Software Guard Extensions} (SGX). The SGX remote attestation techniques and the contained execution features that this method provides, can register each entity of the system as a trusted node. The capabilities that the SGX provides can secure the MSP in all the phases that it is invoked; including each node's registration, the transaction's signature and its verification. An SGX enabled MSP can also mitigate various privacy risks in Fabric and enhance the network's defense  against possible attacks. Therefore, the attack surface of the membership service is required to be analyzed completely and rigorous proofs should to be created to formally quantify all the risks that are aligned with it.

\subsubsection{Pseudonymizing the endorsers}
\label{Pseudonymizing the Endorsers}
Motivated by the aforementioned  risks regarding the endorsers identification, the authors in \cite{Mazumdar}, created a  ring signature scheme, named \textit{Fabric’s Constant-Sized Linkable Ring Signature} (FCsLRS) to pseudoanonymize the endorsers' identities. They implemented this  signature scheme in GO and provided experimental analysis about its security and performance by shifting the \textit{Rivest-Shamir-Adleman} (RSA) modulus size. In the anonymized   endorsement system that they have implemented,  a   threshold  endorsement  policy  needs a set of endorsers to approve a transaction without revealing their identity and  only by counting and checking individual valid ring signatures. A similar work of \cite{Andola}, evaluated the outcome  of a DoS attack on the endorsers and proposed two anonymization techniques; the first by using \textit{verifiable random functions} (VRFs) and the second by using pseudonyms. In both cases, there is a trade-off between the  network's efficiency   and its security; with both the sender and the receiver of a proposed transaction in a private channel to be anonymized \cite{Andola}. However, this  approach certified  with security proofs of \textit{Signature Unforgeability and Unlinkability in Ciphertext} (UN-C) that Fabric can be immune to DoS  and wormhole attacks. 

\section{Privacy Preservation}
\label{Privacy Preservation}

With the current emerge of the blockchain technology, the transactional data are being made  credible (via consensus) and shareable (via the distributed ledger). However, the adoption of this advancement endangers the disclosure of the users' or the companies' private data. Actually, no user would want its private information to be revealed to unauthorized entities exterior to the network and no company would want its competitors to know any private information regarding the costs, the prices or the annual salaries of its employees. Therefore, the collection of the sensitive information   and its secure storage  to the distributed ledger is a crucial issue, since it has to be complied with the  \textit{General Data Protection Regulation} (GDPR). Therefore, in this section, we present Fabric's privacy preservation mechanisms and the few but critical advancements that are needed to be made in order to further enhance the clients (users' or companies') privacy.

\subsection{Privacy techniques and Fabric's specification}
\label{Privacy techniques: Fabric's specification}
Fabric supports significant privacy protection mechanisms. Starting from its  permissioned nature that authorizes the participating entities of the network to strongly authenticate their identity several features are provided to  accommodate the necessary flavors of privacy.

\subsubsection{Channels}
A channel is a state partition with its own access policy rules and transactions' ordering  mechanism. Each channel is managed by a set of peers and it is associated with some policies that provide access to the corresponding resources (such as, the ledger's state, the transactions included in it and the corresponding chaincodes). When a peer registers to a channel, which is characterized by a unique identifier; then the corresponding ledger is created and run on this peer allowing it to manage an identical and consistent data store with the rest of the channel's peers. Therefore,  privacy-preservation mechanisms such as the channels, are highly important in cases of providing blockchain solutions into a consortium environment (where the consortium is comprised by a number of organizations or parties with common business goals). 
 
\subsubsection{Private data collection}
While the channels are devoted to the preservation of the information's privacy, by allowing the information to be stored separately; the \textit{Private Data Collection} (PDC) can preserve  the privacy of data  from another perspective \cite{Ma}. The DPC allows, only a stated set of peers in a channel to  preserve the actual data, while the remaining peers access only its existence proof. The PDC is created to provide to the peers the capabilities of endorsing, committing, or queering private data without being forced to create a new channel and add additional overhead. 

The PDC is actually an accumulation of the following elements: $1)$ the actual private data; which are sent from/to authorized peers by means of a gossip protocol, stored on the peer's private state databases and can been seen only by this set of nodes and not by the OSNs (which are not involved in this case), and $2)$ the hash of private data that is executed, ordered, and stored on each peers ledger as evidence of the existence of the transaction.  In some use cases, where a peer of a PDC wishes to share private data with other peers – for example, to transfer any asset to a \textit{trusted third party} (TTP), the TTP can produce the hash of the private data and subsequently examine if the output of the hash value is consistent with the hash that is stored on the channel's ledger and thus, prove the existence of the transaction. 
 
The ``right-to-be-forgotten'' can be used  in the private transactions, since each peer can erase its own private database at any time, with the  data  itself  to be  deleted irreversibly and the hash  pointing  to  the  underlying  private information to  still  exist. An  additional  aspect that is  implemented  with the private  transactions  is  the  limitation  of  usage.  A  ``BlockToLive''  policy can be defined  for  each  PDC to determine an amount  of  time that has to be  elapsed with  the  concealed  private database  to be  automatically  erased. 
 
Despite the privacy that PDCs provide, they should be used with caution, since the metadata of  the private information, is much more than metadata and can be used to unlock the real private data. In this attack scenario, the unauthorized peers of the same channel  can observe the   shared ledger and detect if the private transactions occur periodically.  

Concluding in this section, the obtained outcomes are that some peers will have full access to the ledger and others may only see what they are allowed to. In the case, where the transactional data must remain hidden during   ordering from some peers of the  same channel and the OSNs, the implementation of PDCs is the solution.

\subsubsection{Zero-Knowledge Proofs (ZKPs)}
ZKPs establish a significant cryptographic primitive to preserve and improve privacy in the blockchain platforms. In Fabric's case, there are two privacy preserving mechanisms that are achieved  with the implementation of ZKPs, and are:

\begin{itemize}
\item Identity mixer \cite{privacyandroulaki} that leverages ZKPs to provide to the clients  anonymous authentication regarding their transactions' proposals. The implementation of ZKPs might have a significant impact when a client's  actual identity and the attributes that it is associated with, must be kept secret from the network (such as, the peers). For example, if the  peers wish to verify that a transaction is indeed sent from a correct client, which is either a  member of a specific organization; (referred in \cite{privacyandroulaki} as  ``membership proof''), or it indeed possesses a specific set of attributes (also referred as ``selective disclosure of attributes''). In both cases the identity mixer verifies that the the client’s identity is not disclosed.

\item \textit{Zero-Knowledge Asset Transfer} (ZKAT) \cite{privacyandroulaki}  is a method built on top of anonymous authentication mechanisms provided by the identity mixer that also uses  ZKPs in various applications aiming at asset management along with audit support (referred as \textit{Zero-Knowledge Asset Transfer} (ZKAT)). With this privacy-preserving mechanism, the clients can issue  transactions without disclosing any other information to the peers regarding the exchange of assets; but only the evidence that each transfer is complied with the  asset management rules. 
\end{itemize}

\subsection{Privacy-preservation challenges and open issues}
\label{Privacy's Challenges  -  Open  Issue}

The aforementioned analysis indicates that appropriate safeguards are in place to mitigate privacy risks in the context of Fabric. Although Fabric offers strong privacy-preserving mechanisms to meet  the necessary flavors of privacy, the research in this field is still ongoing. Thus, some other possible approaches that could be appropriately adopted in the (near) future are: 

\subsubsection{Non-interactive ZKPs}
Although Fabric prevents unauthorized peers to access channel resources, the transactional data is disclosed to all the channel peers. This limitation can be overcome with FabZK \cite{Kang}.  FabZK is a proposed extension for Fabric  to support auditable privacy-preserving chaincode by means of verifiable and well-structured   cryptographic primitives, pertaining \textit{non-interactive Zero-Knowledge Proofs} (NIZKPs)  on Pedersen commitments. The proposed protocol, manages a set of APIs for the client code and  the chaincode to establish  automated  validation, while it improves the transactions' performance with  two validation steps, in which each party executes active  and  lightweight  auto-validation. FabZK showed significant results in \cite{Kang}, with its cryptographic  primitives to outperform other approaches (such as the zkLedger \cite{zkLedger}) when the  NIZKPs are generated  and  verified. 

\subsubsection{Post-quantum signatures}
To ensure secure communication, Fabric relies on \textit{Public Key Infrastructure} (PKI) for the digital signatures and the digital  identities   that  are perilous to the operational security of its network. Moreover, the GDPR demands “consistent” methods to be employed and protect each user's personal identifiable information. However, Fabric's ecosystem is not post-quantum secure, making all the information that is disseminated over the network to be vulnerable to malicious decryption  techniques by a large scale of quantum computers. Therefore, it is left to see, if post-quantum digital signatures are going to be implemented in the (near) future  \cite{Transitioning}.

\section{Conclusions}
\label{sec.concl}
Appropriate implementations and counter-measures to mitigate possible risks in Hyperledger Fabric are discussed in this paper. 

From the consensus protocols security perspective, BFT consensus protocols, such as BFT-SMaRt are currently being researched, since they can provide a significant transaction throughput and tolerance against malicious OSNs. Moreover, BFT protocols have not yet been deployed in production environments; it remains to be seen whether and how such protocols are going to be adopted by the Hyperledger Project. 

Regarding the smart contracts, a plethora of techniques have been discussed to harden the chaincode's security. However, a quite-promising  trend  relies upon  formal verification approaches like theorem proving,  formal modeling and symbolic execution to mitigate the chaincode's risks.  With these approaches, some other aspects of Fabric, such as privacy, performance, and scalability can  be analyzed thoroughly.

From the network's security perspective, representative attacks were discussed against the network when the MSP is compromised and proactive solutions having been proposed as  possible mitigation methods. An alternative/complementary option would be to employ a TEE, such as Intel’s SGX, to address insider threats and also DDoS attacks resulting from the manipulation or aversion of the chaincode's execution.  Techniques to mitigate wormhole attacks vary from those relying on the anonymization of the senders and recipients in the transactions inside a channel, to those employing group signature approaches. Due to the need for accountability in Fabric, such solutions need to be further assessed possibly along with privacy preserving solutions.  
 
Appropriate safeguards to implement the basic privacy requirements, are discussed to mitigate the privacy risks in the context of Fabric.  The implementation of ZKPs can achieve anonymous client authentication with identity mixer,  privacy-preserving exchange of assets with ZKAT, and the  “right to be forgotten”  can be efficiently implemented with PDCs. As the research in this area is rapidly evolving, other promising approaches have been identified that could be adopted in the (near) future, such as the implementation of NIZKP and post-quantum digital signatures. These two areas seem to be a quite promising for research regarding the privacy preservation mechanisms of Fabric.

\bibliographystyle{IEEEtran}
\bibliography{Paper}
\end{document}